\documentclass[preprint,review,12pt]{elsarticle}

\usepackage{lineno}
\usepackage{hyperref}
\modulolinenumbers[5]
\usepackage{setspace}

\usepackage{float}
\restylefloat{table}

\usepackage{graphics}
\usepackage[fleqn]{amsmath}
\usepackage{amssymb}
\usepackage{amsthm}

\usepackage{algorithm}
\usepackage{algorithmic}

\usepackage{listings}

\usepackage{multirow}
\usepackage{makecell}
\usepackage[usenames]{xcolor}
\usepackage{enumitem}
\usepackage{soul}

\journal{Elsevier}

\begin{document}

\begin{frontmatter}

\title{XAMG: A library for solving linear systems with multiple right-hand side vectors}

\author{B. Krasnopolsky \corref{author1}}
\author{A. Medvedev \corref{author2}}

\cortext[author1] {\textit{E-mail address:} krasnopolsky@imec.msu.ru} 
\cortext[author2] {\textit{E-mail address:} a.medvedev@imec.msu.ru} 
\address{Institute of Mechanics, Lomonosov Moscow State University, 119192 Moscow, Michurinsky
ave.~1, Russia}

\begin{abstract}	
This paper presents the XAMG library for solving large sparse systems of linear algebraic equations
with multiple right-hand side vectors. The library specializes but is not limited to the solution of
linear systems obtained from the discretization of elliptic differential equations. A corresponding 
set of numerical methods includes Krylov subspace, algebraic multigrid, Jacobi, Gauss-Seidel, 
and Chebyshev iterative methods. The parallelization is implemented with MPI+POSIX shared memory 
hybrid programming model, 
which introduces a three-level hierarchical decomposition using the corresponding per-level
synchronization and communication primitives. The code contains a number of optimizations, including
the multilevel data segmentation, compression of indices, mixed-precision floating-point calculations,
vector status flags, and others. {The XAMG library uses the program code of the
well-known \textit{hypre} library to construct the multigrid matrix hierarchy. The XAMG's own
implementation for the solve phase of the iterative methods provides up to a twofold speedup
compared to \textit{hypre} for the tests performed. Additionally, XAMG provides extended
functionality to solve systems with multiple right-hand side vectors.}
\end{abstract}


\end{frontmatter}

\begin{table*}[!hbt]
\begin{tabular}{|l|p{6.5cm}|p{6.5cm}|}
\hline
\textbf{Nr.} & \textbf{Code metadata} &  \\
\hline
C1 & Current code version & Version 1.0 \\
\hline
C2 & Permanent link to code/repository used for this code version & $https://gitlab.com/xamg/xamg$ \\
\hline
C3 & Code Ocean compute capsule & N/A \\
\hline
C4 & Legal Code License   & GPLv3 \\
\hline
C5 & Code versioning system used & git \\
\hline
C6 & Software code languages, tools, and services used & C++, MPI \\
\hline
C7 & Compilation requirements, operating environments \& dependencies & C++11 compiler, POSIX, MPI,
\textit{hypre}. The build instructions: $https://gitlab.com/xamg/xamg/-/wikis/docs/XAMG\_build\_guideline$ \\
\hline
C8 & If available Link to developer documentation/manual & \\
\hline
C9 & Support email for questions & Gitlab issue tracker section \\
\hline
\end{tabular}
\caption{Code metadata}
\label{} 
\end{table*}


\linenumbers

\section{Motivation and significance}
\label{}
Despite the wide use of open-source libraries for solving systems of linear algebraic equations in
computational software, in some cases there is still the need for developing novel libraries, which can
extend the functionality of well-known ones or add some features, allowing to improve
performance of the calculations. The \textit{hypre} library~\cite{hypre} is an example of an outstanding piece of
software containing powerful and highly scalable algorithms for solving large sparse systems of
linear algebraic equations~(SLAEs). The classical algebraic multigrid method~\cite{Trottenberg}
implemented in \textit{hypre} is a robust method widely applied for solving SLAEs derived from elliptic or
parabolic differential equations. However, the implementation of the methods in \textit{hypre}, as well as
many other libraries containing various modifications of multigrid methods,
is limited by algorithms performing calculations with a single right-hand side vector (RHS).
Meanwhile, the solution of systems with multiple RHSs,
$$A \, X = B,$$ 
where $A \in \mathbb{R}^{n \times n}$, and $X,B \in \mathbb{R}^{n \times m}$, $m \ll n$, in terms of
computational efficiency can be preferable compared to multiple solutions of systems with single RHS.
The reasons to this include: (i)~increasing the arithmetic intensity, (ii)~regularization of the memory access
pattern, (iii)~vectorization of calculations, and others.

The SLAEs with multiple RHSs occur in a variety of mathematical physics applications. For
example, the need for solving the corresponding systems of linear algebraic equations appears in
structural analysis applications~\cite{Feng1995}, in uncertainty quantification
problems~\cite{Kalantzis2018}, in Brownian dynamics simulations~\cite{Liu2012}, quantum
chromodynamics~\cite{Clark2018}, {computational fluid
dynamics~\cite{KrasnopolskyCPC2018},} and others.

Despite the variety of applications and several advantages mentioned above, implementations of
iterative methods for solving SLAEs with sparse matrices and multiple RHSs can rarely be
found in publicly available libraries of numerical methods. {Among the only few
exceptions is the Trilinos library~\cite{Trilinos} containing the implementation of several Krylov subspace and
aggregation-based algebraic multigrid methods. These, however, do not fully cover the entire
set of the methods used in numerical modeling, e.g. the classical algebraic
multigrid methods, provided by the \textit{hypre} library. This issue complicates the development of the} computational
algorithms performing calculations with multiple RHSs.

The XAMG library focuses on the solution of a series of systems of linear
algebraic equations with multiple RHSs (that, e.g. occur in incompressible turbulent flow
simulations~\cite{KrasnopolskyCPC2018}). The code extends the functionality of the classical
algebraic multigrid method, implemented in \textit{hypre}, and provides implementations of both classical and
merged formulations of the Krylov subspace iterative methods~\cite{Krasnopolsky_CAMWA2020} to solve
systems with multiple RHSs. The XAMG library aims to develop an optimized implementation for the
solve phase of the methods, while the \textit{hypre} library can be reused ``as is'' to construct
a multigrid matrix hierarchy.

The library provides additional optimization features, including the hierarchical multilevel
parallelization, low-level intra-node shared memory communications featuring synchronization
primitives based on POSIX shared memory and atomics, multi-block data segmentation in accordance
with hierarchical parallelization, mixed-precision floating-point calculations, compression of
indices, vector status flags, and others, which allow to outperform the \textit{hypre} library calculation
times.

\section{Software description}
\label{}

\subsection{Mathematical methods}
\label{lab:meth}
The XAMG library contains a set of numerical methods typically used for solving systems of linear
algebraic equations coming from the discretization of elliptic partial differential equations. These
include several Krylov subspace iterative methods (CG~\cite{Hestenes1952} and
BiCGStab~\cite{Vorst1992}), classical algebraic multigrid method~\cite{Trottenberg}, block Jacobi
and symmetric Gauss-Seidel methods, Chebyshev iterative method, and a direct solver. All these methods
are adopted to solve SLAEs with multiple RHSs. In addition to the classical BiCGStab method, the
library also contains several modified methods, including the Reordered BiCGStab~\cite{RBiCGStab},
Pipelined BiCGStab~\cite{Cools2017} and merged formulations of these
methods~\cite{Krasnopolsky_CAMWA2020}, which combine vector updates and dot products to minimize the
overall amount of vector read/write operations. The possible combinations of the methods, which can
be used as a standalone solver, preconditioner, or smoother with the multigrid method, are
summarized in Tab.~\ref{tab:methods_usage}. The full list of the methods and corresponding method
parameters is presented in~\cite{xamg_params_ref}.

\begin{table*}[!th]
  \caption{Combinations of the methods and solver types implemented in the XAMG library.}
  \label{tab:methods_usage}
\centering
\begin{tabular}{ | c | c | c | c | }
\hline
	Method & Solver & Preconditioner & Pre-/post-smoother \\
\hline
	CG        & + & - & - \\
\hline
	BiCGStab  & + & + & + \\
\hline
	MultiGrid & + & + & - \\
\hline
	Jacobi & + & + & + \\ 
\hline
    Gauss-Seidel & + & + & + \\
\hline
	Chebyshev & - & + & + \\
\hline
	Direct    & + & - & - \\
\hline
\end{tabular}
\end{table*}

\subsection{Advantages of calculations with multiple RHSs}
\label{lab:advantages}
The solution of SLAEs with multiple RHSs opens up opportunities for several optimizations.
The main advantage of using iterative methods with performing simultaneous independent solutions for
multiple RHSs is an increasing arithmetic intensity of calculations. The flop per
byte ratio is a measure of floating-point operations relative to the amount of memory accesses.
Basic linear algebra operations like vector
updates, dot products, and sparse matrix-vector multiplications (SpMV) are characterized by flop per byte ratio
of only about~0.1. This means that the corresponding operations are memory bound~\cite{Williams2009,
Krasnopolsky_CAMWA2020}, and its performance mostly depends on the
amount of memory traffic and memory bandwidth. The solution of SLAEs with multiple RHSs
allows to load a matrix for SpMV operations with $m$~right-hand sides only once, compared
to $m$~loads when performing $m$ SLAE solutions with a single RHS. This typically
leads to {about a twofold} reduction in the total amount of memory traffic for SLAE
solver per each RHS, an increase of the flop per byte ratio, and, subsequently, to a significant calculation speedup.

The use of specialized data storage formats for sparse matrices is necessary for performing the
operations with large matrices arising in numerical simulations. The compressed sparse row~(CSR)
format~\cite{Saad2003} is the universal one, which is widely used for the matrices of general form.
However, the implementation of SpMV operation for CSR format requires indirect memory accesses,
which reduces cache efficiency and gives no chance for loop vectorization. The generalized SpMV
operation performed for multiple RHSs allows to partially regularize an access to the memory
and vectorize the computations over the right-hand sides.

\subsection{Software architecture}
\label{sec:arch}

The XAMG library is designed as a header-only template C++ library. The template polymorphism is
used to implement variability of all basic data types for matrices and vectors, both interface-level
and internal ones. The number of RHSs is also given as a template parameter.

The specialization of the number of RHSs at compile time is a key design point.
This approach allows to vectorize all the relevant subroutines with respect to a constant
range of loops iterating over the RHSs. Keeping in mind the appropriate vectors' data
placement and some pragma directives code instrumentation, such a design allows a regular
C++ compiler to vectorize subroutines to the greatest extent possible.

\begin{figure*}[t!]
\centering
\includegraphics[width=13.0cm]{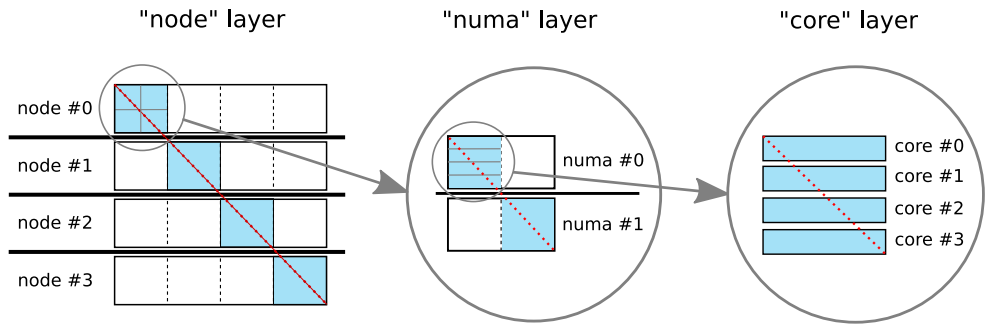}
\caption{The hierarchical matrix decomposition.}
\label{fig:hierarchy}
\end{figure*}

One can highlight the following XAMG program code elements:
\begin{itemize}[noitemsep]
\item matrix and vector data structures;
\item basic sparse linear algebra subroutines gathered in ``blas'' and ``blas2'' groups;
\item solver classes inherited from an abstract interface, which exposes the main library
functions: \texttt{setup()} and \texttt{solve()};
\item solver parameter classes.
\end{itemize}
The matrix and vector data structures are used to store the corresponding data objects.
The ``blas'' group of subroutines consists of a collection of primitives operating with vectors
(linear updates, dot products, etc.) The ``blas2'' group of subroutines includes several SpMV-like
matrix-vector operations. Solver classes implement various iterative methods used to solve systems
of linear algebraic equations. Finally, the solver parameter classes are used to store the lists of
numerical method parameters, specifying each numerical method's configuration.

The \verb|setup()| virtual function of the MultiGrid solver class implements the
classical algebraic multigrid ``setup'' phase. This is done by wrapping up the \textit{hypre}
functions to construct the multigrid matrix hierarchy. This is the only place where the
\textit{hypre} library codebase is used inside the XAMG library program code.

An important design feature of the XAMG library is the hierarchical hybrid parallel programming
model. This model combines the Message Passing Interface (MPI) standard~\cite{MPI_standard} for
cross-node communication and a special kind of shared memory parallel programming model on the
intra-node level, based on POSIX shared memory (ShM) functionality. {The MPI+ShM
model implies that MPI ranks, which reside on the same compute node, allocate and use the common
POSIX shared memory regions to store data objects. Unlike many other MPI+X programming models,
MPI+ShM does not use threading within a node. The MPI rank to CPU core mapping is one-to-one.}

To benefit from better data locality, most well-known distributed linear algebra algorithms
implement a data decomposition that splits the calculations into fully local and fully non-local
parts~\cite{Gorobets2018}.
The MPI+ShM programming model also implements a specific data decomposition for matrix and vector
data structures. The idea behind this decomposition is the hierarchical segmentation of the data
into: (i) those with only local algorithmic dependencies, and (ii) those requiring some parallel
communication to work with. In Figure~\ref{fig:hierarchy}, the blue-colored boxes represent the
``local dependency'' blocks, whereas the white-colored ones are for ``remote dependency''. The local
dependency blocks can also be called ``diagonal blocks'' since they contain matrix diagonal 
elements for square matrices. The Figure~\ref{fig:hierarchy} also illustrates why the data
decomposition is called hierarchical. In fact, the recursive data decomposition introduces three
logical layers: the node layer, the numa layer, and the core layer. They are meant to reflect the
natural hierarchy of a modern CPU-based high performance computing system hardware, consisting of
the compute nodes, the NUMA blocks inside each node, and the CPU cores.

The hierarchical decomposition principle makes it possible to split the local and non-local data
dependencies. As a practical result of this, the number of non-local MPI communications
(on a cross-node layer) as well as non-local data accesses (for a cross-NUMA layer) is reduced
significantly.

To summarize, the hybrid MPI+ShM parallel programming model implies: (i)~the direct shared memory
data access instead of MPI communications on an intra-node scope, and (ii)~a specific hierarchical data
and algorithms decomposition. Both features combined significantly improve the single-node parallel
performance and the multi-node scalability compared to the pure MPI model.

{The data compression features of the XAMG library try to benefit from reducing
the number of data transfers in the main library algorithms and communication procedures. These features
include the turnover to the reduced precision for floating-point numbers and using lesser bits for integer
indices for the matrices stored in the CSR format. The reduced precision floating-point numbers
typically cannot be used for the whole solver without losing the solution accuracy. However, it is
known that switching to the reduced precision when performing preconditioner
calculations~\cite{Buttari2008, Sumiyoshi2014}
may be a compromise variant, preserving the basic precision of the resulting vector and having only a
minor or no reduction of the convergence rate. Depending on the fraction of the calculations with
the reduced precision in the overall SLAE solver calculations, the 5-30\% cut on the full amount of
data to transfer allows to obtain the proportional calculations speedup. Since the reduced
precision feature is only applied to a part of the multigrid matrices hierarchy, this data
compression model is called ``mixed-precision''.}

{The second type of optimizations related to compression of indices for the CSR
matrix storage format is applicable to all matrices. The hierarchical data decomposition leads to a
multiblock matrices representation, and for the small blocks CSR indices data types can be changed
to the smaller ones like uint16\_t or uint8\_t. This modification also results in reducing the
amount of memory accesses.}

\subsection{Software usage examples}
\label{}

The main API of the XAMG library has a header-only C++11 template-based interface, but the C
language binding also exists. The below overview shows an example of using the main C++ library
interface.

\begin{figure*}[t!]
\begin{lstlisting}[language=C++]
  1 #include <xamg/xamg_headers.h>
  2 #include <xamg/xamg_types.h>
  3 #include <xamg/init.h>
  4 #include <xamg/blas/blas.h>
  5 #include <xamg/blas2/blas2.h>
  6 #include <xamg/solvers/solver.h>
  7
  8 int main(int argc, char *argv[]) {
  9     using FP = double;
 10     using matrix_t = XAMG::matrix::csr_matrix<FP, uint32_t, uint32_t,
 11                                                   uint32_t, uint32_t>;
 12     constexpr uint16_t NV = 1;
 13
 14     XAMG::init(argc, argv, "nnumas=1:ncores=1");
 15
 16     matrix_t csr_file_mtx;
 17     XAMG::vector::vector csr_file_x, csr_file_b;
 18     XAMG::io::read_system<matrix_t, NV>(csr_file_mtx, csr_file_x,
 19                                         csr_file_b, "matrix.csr");
 20
 21     XAMG::matrix::matrix A(XAMG::mem::DISTRIBUTED);
 22     XAMG::vector::vector x(XAMG::mem::DISTRIBUTED);
 23     XAMG::vector::vector b(XAMG::mem::DISTRIBUTED);
 24     auto part = XAMG::part::make_partitioner(csr_file_mtx.nrows);
 25     XAMG::matrix::construct_distributed(part, csr_file_mtx, A);
 26     XAMG::vector::construct_distributed<FP, NV>(part, csr_file_x, x);
 27     XAMG::vector::construct_distributed<FP, NV>(part, csr_file_b, b);
 28
 28
 29     XAMG::params::global_param_list params;
 30     params.add("solver", {"method", "PBiCGStab"});
 31     params.add("preconditioner", {"method", "Jacobi"});
 32     params.set_defaults();
 33
 34     auto solver =
 35       XAMG::solver::construct_solver_hierarchy<FP, NV>(params, A, x, b);
 36
 37     solver->solve();
 38
 39     XAMG::finalize();
 40     return 0;
 41 }
\end{lstlisting}
\vspace{-0.5cm}
\caption{The XAMG library general usage example, C++ API.} \label{fig:generalusage}
\end{figure*}

The solver code is configured at compile-time depending on the classes' and functions' concrete
template parameters. The C++ template instantiation is a mechanism behind this. A basic matrix
type is instantiated with integer types (uint8\_t, uint16\_t, and uint32\_t are allowed) and
floating-point type (float or double). The integer template types specify the storage types for
nonzero matrix elements indices. The floating-point template type defines the storage type for
the values of these elements. The same floating-point type is also used to instantiate the functions
operating with input and output vectors. The majority of the XAMG data structures and library
functions have an additional template parameter $NV$\footnote{The number of RHSs, $m$,
is denoted in the source code as a template parameter $NV$.}, the number of RHSs within
a single SLAE solution operation.

The code listing in Figure~\ref{fig:generalusage} presents a basic working code portion that uses
the XAMG library to solve a linear system with a sparse matrix stored in the CSR format.
Lines 9-12 set up the compile time definitions for the above mentioned basic data types and the
number of RHSs for a solution. The library initialization call in line~14 passes an optional
colon-separated configuration string. This string can be used to specify the
parameters of the hybrid three-level parallel programming model. The details on these parameters
are given in the program documentation.

Lines 16-19 represent a basic code to read a matrix from a binary file. The file is supposed to
hold a CSR format matrix, an RHS, and an initial guess vector (as an option). 
Lines 21-27 fill in the distributed versions of the matrix and the vectors with a correct 
decomposition, and map the data structures to
the MPI+ShM data hierarchy. The desired solver parameters are set in lines 29-32. A preconditioned
BiCGStab method is set for usage in this example, with the Jacobi method serving as a
preconditioner. All the default parameters for the solver/preconditioner pair are set up in line 32.
Lines 34-35 create a C++ object representing a solution plan based on the solver and preconditioner
parameters, and this plan becomes connected to the distributed matrix and the vectors $A$, $x$, and
$b$. The SLAE solution process for $NV$ RHSs is done in line 37. The solution result is
stored in $x$ vector, and some solution statistics are held in the $stats$ variable of the $solver$
object.

\begin{figure*}[t!]
\begin{lstlisting}[language=C++]
  1 XAMG::params::global_param_list params;
  2 params.add("solver", {"method", "PBiCGStab"});
  3 params.add_map("solver", {{"max_iters", "20"}});
  4 params.add_map("preconditioner", {{"method", "MultiGrid"},
  5                                   {"max_iters", "1"},
  6                                   {"mg_agg_num_levels", "2"},
  7                                   {"mg_coarse_matrix_size", "500"},
  8                                   {"mg_num_paths", "2"}});
  9 params.add_map("pre_smoother", {{"method", "Chebyshev"},
 10                                 {"polynomial_order", "2"}});
 11 params.add_map("post_smoother", {{"method", "Chebyshev"},
 12                                  {"polynomial_order", "2"}});
 13 params.set_defaults();
\end{lstlisting}
\vspace{-0.5cm}
\caption{An example of the numerical method parameters configuration with the BiCGStab method,
supplemented with the algebraic multigrid preconditioner.} \label{fig:mgparams_src}
\end{figure*}

The example set of parameters for a simple algebraic multigrid configuration of a solver is shown in
Figure~\ref{fig:mgparams_src}. In lines 4-8, the multigrid method is set as a preconditioner for a
preconditioned BiCGStab solver. A few parameters which tune the multigrid hierarchy setup process
are set explicitly. In lines 9-12, we take care of an explicit smoothers set up for the
multigrid method. Specifically, the Chebyshev method with a polynomial order of~2 is set. The
\verb|set_defaults()| function call in line~13 ensures that all the solver parameters, as well as
preconditioner and smoother methods, which are not set explicitly in previous lines, are correctly
initialized with the default values.

\section{Example case: performance evaluation}
\label{sec:example_case}

The example case presents the study on the efficiency of the XAMG library and the impact of
implemented optimization features on the overall code performance. {It demonstrates
(i)~the XAMG library performance compared to \textit{hypre}, (ii)~the effect of data compression
optimizations on the calculation time reduction, (iii)~the performance gain due to calculations with
multiple RHSs, and (iv)~the parallel efficiency results for various numbers of RHSs}. The
Lomonosov-2 supercomputer is used for the performance evaluation to demonstrate the potential of the
developed library. It consists of compute nodes with a single Intel Xeon Gold~6126 processor and
InfiniBand FDR interconnect. In all the cases all available 12~physical CPU cores per node are
utilized during the calculations. {The XAMG library is compiled with the Intel
C/C++ compiler~2019\footnote{The compilation options are: \texttt{-ip -O3 -no-prec-div -static
-fp-model fast=2 -xCORE-AVX512}} and Intel MPI Library, Version~2019 Update~9.} The internal
integration test application \verb|xamg_test| is used for the tests; the YAML configuration files
are provided within the XAMG source code repository~\cite{xamg_repo} in \verb|examples/test/yaml/|
directory.


The test runs are performed for several test SLAEs. They correspond to the systems obtained as a
result of spatial discretization of the 3D Poisson's equation $$\Delta u = f$$ with the 7-point
stencil in various computational domains, including the regular grid in the cube (grids with $150^3$
and $250^3$~cells), and the computational grid, corresponding to the direct numerical simulation of
the incompressible turbulent flow in a channel with a matrix of wall-mounted cubes (matrix of 9.7~mln.
unknowns~\cite{KrasnopolskyCPC2018}). The corresponding matrix generators are embedded into the
\verb|xamg_test| integration test application.

\subsection{Calculations with single RHS}
\label{sec:1rhs}
The first set of experiments is performed with two test SLAEs corresponding to cubic computational
domain with the grids of $150^3$ and $250^3$ unknowns. The numerical method configuration, identical
for both XAMG and \textit{hypre}, comprises the BiCGStab method supplemented with an algebraic multigrid
preconditioner and hybrid symmetric Gauss-Seidel smoother. The runs performed include five
different test series:
the \textit{hypre} library tests\footnote{The most recent \textit{hypre} library version~2.20.0
available to date is used in the performance comparison tests.} in pure MPI and hybrid MPI+OpenMP execution
modes, the XAMG library tests in pure MPI and MPI+ShM execution modes, and the XAMG library
tests in hybrid MPI+ShM mode with data compression optimizations.

\begin{figure*}[t!]
\centering
\includegraphics[width=6.0cm]{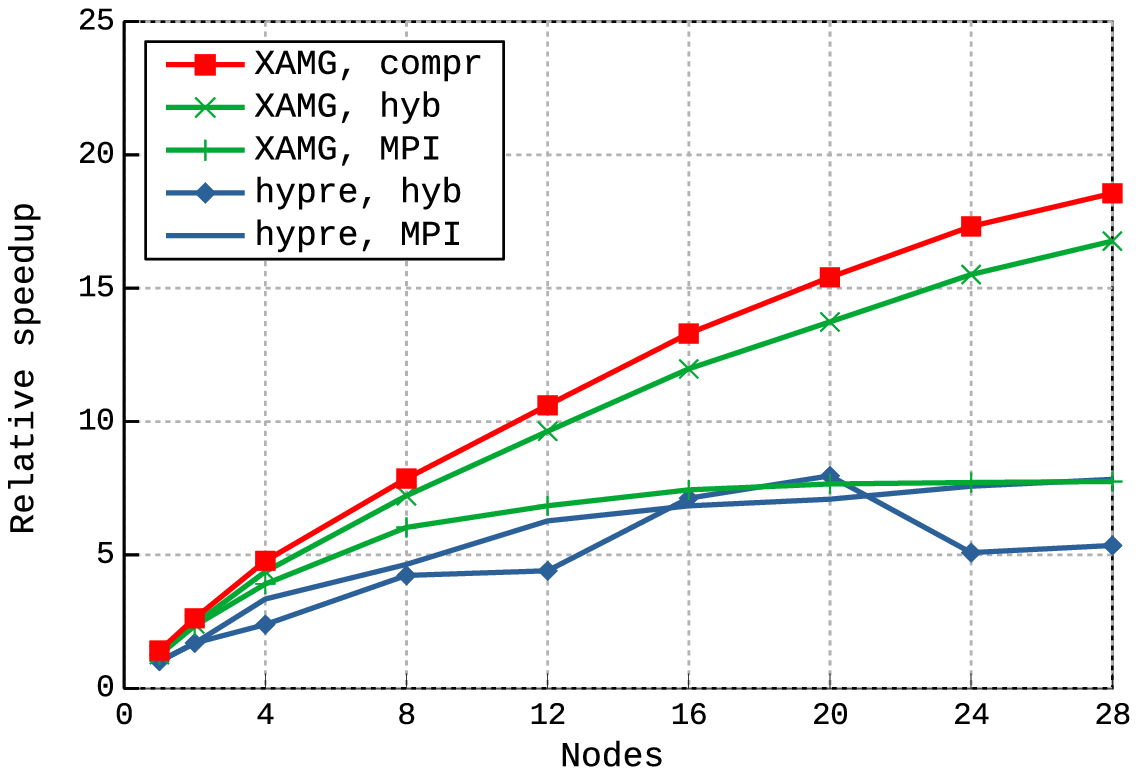}
\includegraphics[width=6.0cm]{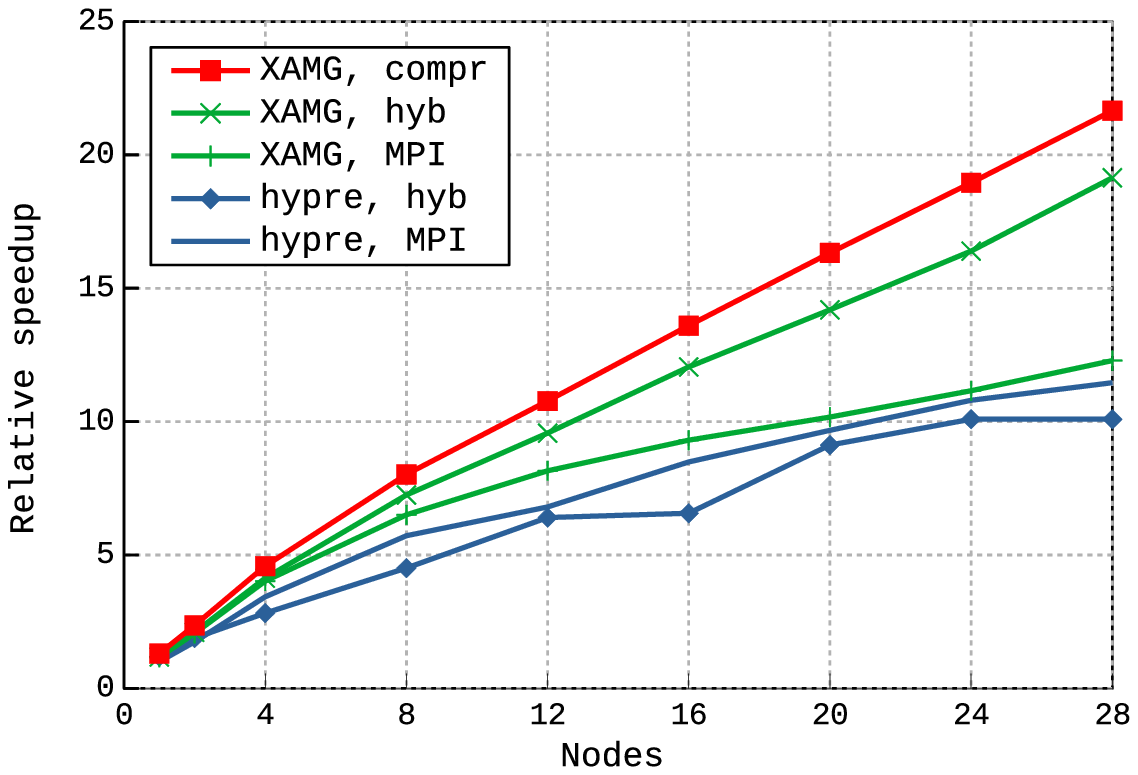}
\caption{Relative calculation speedup for the XAMG and \textit{hypre} libraries in various execution modes,
test matrices with $150^3$ (left) and $250^3$ (right) unknowns.}
\label{fig:scal}
\end{figure*}

The obtained performance evaluation results are shown in Figure~\ref{fig:scal}. The data is
presented in terms of the relative speedup, which is defined as a ratio of the single node
calculation time for the \textit{hypre} library in a pure MPI
mode, to the calculation time with the specific number of compute nodes and execution mode:
$$ 
S^i_p = \frac{T_1^{hypre, MPI}}{T_p^i}.
$$ 
Results presented in Figure~\ref{fig:scal} show similar scalability
for the pure MPI implementations of XAMG and \textit{hypre}.
However, the MPI+ShM hybrid programming model of the XAMG library provides a big scalability
improvement: with MPI+ShM programming model enabled, XAMG outperforms
\textit{hypre}{, employing the MPI+OpenMP model,} more than twice.
Additionally, the data compression optimization of XAMG adds an extra 10\% speedup.

{The difference in the hybrid programming model benefits, observed 
when comparing XAMG and \textit{hypre} libraries, is related to different
hybridization principles for these codes. The MPI+OpenMP hybrid programming model
is known to suffer from fundamental multithreading issues within the 
MPI parallel paradigm~\cite{Amer2015}. The MPI+ShM programming model 
avoids these pitfalls with the single-threaded design. Among the other reasons can be the difference
in data decomposition principles, solver implementation aspects, implementation of inter-node
communications, or some other algorithmic or programming aspects. However, this topic cannot be elaborated 
in detail within the scope of this article.}

\subsection{Calculations with multiple RHSs}

\subsubsection{Single-node performance}
\label{sec:mrhs_1node}

{
The performance gain parameter is a key characteristic indicating the benefit of solving the SLAE
with multiple RHSs over multiple solutions with a single RHS. This parameter is defined as:
$$
P_m = \frac{m T_1}{T_m},
$$
where $m$ is the number of RHSs, $T_1$ is the calculation time for a single RHS, and $T_m$ is the
calculation time for $m$ RHSs. The number of RHSs in the tests varied in the range 1--64 for the
smaller test matrix, and in the range 1--16 for the bigger ones (due to the single node memory
capacity limitations). The corresponding results are presented in Figure~\ref{fig:scal_mrhs}. The
plot shows at least twofold performance gain when performing calculations with
multiple RHSs, which is in agreement with the theoretical estimates~\cite{KrasnopolskyCPC2018}. The
peak values vary in the range 2-2.5, depending on the fraction of SpMV-like operations
in the overall SLAE solver calculation loop.}

\begin{figure}[t!]
\centering
\includegraphics[width=6.0cm]{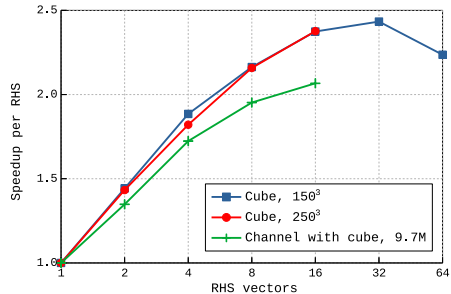}
\caption{Performance gain for the calculations with multiple RHSs.}
\label{fig:scal_mrhs}
\end{figure}

\subsubsection{Parallel efficiency}

The XAMG library performance is also evaluated by the multiple RHS calculations parallel
efficiency. The corresponding tests are performed with the test matrix of 9.7~mln. unknowns with 1, 4, and 16~RHS
vectors. The parallel efficiency results, 
$$
E_p = \frac{T_1}{p \, T_p},
$$ 
are obtained for the hybrid execution mode. The corresponding results are presented in
Figure~\ref{fig:scal_mrhs}. These results show that the parallel efficiency for multiple RHSs
is, at least, not worse compared to the one for the single RHS calculations.

\begin{figure}[t!]
\centering
\includegraphics[width=6.0cm]{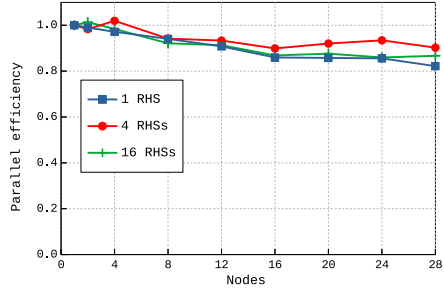}
\caption{The parallel efficiency for the multiple RHSs calculations with the XAMG library, 9.7~mln.
unknowns test matrix.}
\label{fig:scal_mrhs}
\end{figure}


\section{Impact and conclusions}
\label{}

The new linear solver code XAMG is a modern C++ project which adds a rare feature for the libraries
of iterative methods, as it solves systems with multiple right-hand side vectors.
The key advantage of the solver
with multiple RHSs is a potential speedup over multiple solutions with single RHS, and this feature
can be effectively used in some well-known mathematical methods for structural analysis,
uncertainty quantification, Brownian dynamics simulations, quantum chromodynamics, modeling of
incompressible turbulent flows, and others. However, despite evident
advantage in time to solution, this functionality is not implemented in most 
libraries containing robust and scalable iterative methods, like algebraic multigrid.

Among the real applications already taking up this advantage, we highlight an in-house code for
incompressible turbulent flow simulations~\cite{KrasnopolskyCPC2018}, since the XAMG library is
already successfully integrated with it. The integration can be done as well for any other project
written in C/C++ that is able to exploit the multiple RHS feature of a linear solver because the
initial release of the XAMG library has been recently published as an open repository 
on gitlab.com~\cite{xamg_repo}.

The XAMG project also features a number of leading source code and computational method optimizations
which impact the generic problem of a linear solver speedup on modern multicore and manycore
HPC systems. This problem stays aside from the multiple RHS feature. These optimizations like a
novel hybrid parallel programming model (MPI+ShM), mixed-precision calculations, matrix indices
compression, and others make it possible to improve the productivity of a linear solver compared to
other well-known libraries containing robust and scalable iterative methods (e.g. \textit{hypre}). The
modern C++11 codebase allows for broad future research prospects in this direction.

{The experimental performance evaluation shows the XAMG library speedup against
\textit{hypre} by a factor of~2 for the multi-node runs performed. The comparison of calculations
with multiple RHSs against multiple runs with single RHS shows 2-2.5 times speedup for single-node
runs. Additionally, the multi-node parallel efficiency of the multiple RHSs calculations is, at
least, not worse compared to the one for the single RHS.}

\section*{Declaration of competing interest} 
The authors declare that they have no known competing financial interests or personal relationships
that could have appeared to influence the work reported in this paper.


\section*{Acknowledgments}
\label{lab:ack}
The presented work is supported by the RSF grant No. 18-71-10075. The research is carried out using
the equipment of the shared research facilities of HPC computing resources at Lomonosov Moscow State
University.


\bibliographystyle{elsarticle-num}
\bibliography{base_hpc}


%
%
%

\end{document}